# First simultaneous measurement of the γ-ray and neutron emission probabilities in inverse kinematics at a heavy-ion storage ring

M. Sguazzin[1,*], B. Jurado[1,†], J. Pibernat[1], J. A. Swartz[1,‡], M. Grieser[2], J. Glorius[3], Yu. A. Litvinov[3], C. Berthelot[1], B. Włoch[1], J. Adamczewski-Musch[3], P. Alfaurt[1], P. Ascher[1], L. Audouin[4], B. Blank[1], K. Blaum[2], B. Brückner[5], S. Dellmann[5], I. Dillmann[6,7], C. Domingo-Pardo[8], M. Dupuis[9,10], P. Erbacher[5], M. Flayol[1], O. Forstner[3], D. Freire-Fernández[2,11], M. Gerbaux[1], J. Giovinazzo[1], S. Grévy[1], C. J. Griffin[6], A. Gumberidze[3], S. Heil[5], A. Heinz[12], R. Hess[3], D. Kurtulgil[5], N. Kurz[3], G. Leckenby[6,13], S. Litvinov[3], B. Lorentz[3], V. Méot[9,10], J. Michaud[1*], S. Pérard[1], N. Petridis[3], U. Popp[3], D. Ramos[14], R. Reifarth[5,15], M. Roche[1], M.S. Sanjari[3,16], R.S. Sidhu[17,3,2], U. Spillmann[3], M. Steck[3], Th. Stöhlker[3], B. Thomas[1], L. Thulliez[18] and M. Versteegen[1]

[1] Université de Bordeaux, CNRS, LP2I Bordeaux, 33170 Gradignan, France
[2] Max-Planck-Institut für Kernphysik, 69117 Heidelberg, Germany
[3] GSI Helmholtzzentrum für Schwerionenforschung, 64291 Darmstadt, Germany
[4] Université Paris-Saclay, CNRS, IJCLab, 91405 Orsay, France
[5] Goethe University of Frankfurt, 60438 Frankfurt, Germany
[6] TRIUMF, Vancouver, British Columbia, V6T 2A3, Canada
[7] Department of Physics and Astronomy, University of Victoria, Victoria, BC V8P 5C2, Canada
[8] IFIC, CSIC-Universidad de Valencia, 46980 Valencia, Spain
[9] CEA, DAM, DIF, 91297 Arpajon, France
[10] Université Paris-Saclay, CEA, LMCE, 91680 Bruyères-Le-Châtel, France
[11] Ruprecht-Karls-Universität Heidelberg, 69117 Heidelberg, Germany
[12] Chalmers University of Technology, 41296 Gothenburg, Sweden
[13] Department of Physics and Astronomy, University of British Columbia, Vancouver, BC V6T IZI, Canada
[14] GANIL, 14000 Caen, France
[15] Los Alamos National Laboratory, Los Alamos, NM, 87544, USA
[16] Aachen University of Applied Sciences, Aachen, Germany
[17] School of Physics and Astronomy, University of Edinburgh, EH9 SFD Edinburgh, United Kindom
[18] IRFU, CEA, Université Paris-Saclay, 91191 Gif-sur-Yvette, France

*Abstract:* The probabilities for γ-ray and particle emission as a function of the excitation energy of a decaying nucleus are valuable observables for constraining the ingredients of the models that describe the de-excitation of nuclei near the particle emission threshold. These models are essential in nuclear astrophysics and applications. In this work, we have for the first time simultaneously measured the γ-ray and neutron emission probabilities of $^{208}$Pb. The measurement was performed in inverse kinematics at the Experimental Storage Ring (ESR) of the GSI/FAIR facility, where a $^{208}$Pb beam interacted through the $^{208}$Pb(p,p') reaction with a hydrogen gas jet target. Instead of detecting the γ-rays and neutrons emitted by $^{208}$Pb, we detected the heavy beam-like residues produced after γ and neutron emission. These heavy residues were fully separated by a dipole magnet of the ESR and were detected with outstanding efficiencies. The comparison of the measured probabilities with model calculations has allowed us to test and select different descriptions of the γ-ray strength function and the nuclear level density available in the literature.

## I. Introduction

Understanding the de-excitation process of heavy nuclei at excitation energies around the particle emission threshold is essential for the development of nuclear reaction models that predict the cross sections relevant for nuclear astrophysics and applications in nuclear technology. In this excitation-energy range, a heavy nucleus can decay via different competing channels: γ-ray emission, particle emission (e.g. emission of neutrons or protons) and fission. The de-excitation process is ruled by fundamental properties of nuclei such as nuclear level densities (NLD), γ-ray strength functions (GSF), particle transmission coefficients or fission barriers. However, the current nuclear-structure models are unable to predict these properties with sufficient accuracy, which results in significant uncertainties in the calculated cross sections. This is particularly evident in the case of cross sections of neutron-induced reactions of very short-lived nuclei, which are essential for understanding the synthesis of elements via the rapid neutron capture process [1, 2].

---

* Present address: IJCLab, 91405 Orsay, France
† Corresponding author: jurado@cenbg.in2p3.fr
‡ Present address: FRIB, MSU, Michigan 48824, USA

Two-body reactions $X(a,b)Y^*$, such as inelastic scattering and transfer reactions with light projectile nuclei $a$, are well suited to form nuclei $Y^*$ with a wide range of excitation energies, including the particle emission threshold, and to study the de-excitation of $Y^*$ by measuring the probability that $Y^*$ decays via a channel $\chi$ as a function of its excitation energy $E^*$. Possible decay channels are for example γ-ray- ($\chi = \gamma$) and neutron- ($\chi = n$) emission, thus $P_\gamma$ and $P_n$ measure the likelihood that the excited nucleus $Y^*$ decays by the emission of a γ-ray cascade and of a neutron, respectively. The probabilities $P_\chi(E^*)$ are valuable observables to constrain the models describing the fundamental nuclear properties mentioned above. The latter probabilities are determined as follows:

$$P_\chi(E^*) = \frac{N_{c,\chi}(E^*)}{N_s(E^*) \cdot \varepsilon_\chi(E^*)} \qquad (1)$$

where $N_s$ is the number of light ejectiles $b$ measured, the so-called single events. $N_{c,\chi}$ is the number of products of the subsequent decay channel $\chi$ measured in coincidence with the ejectiles $b$ and $\varepsilon_\chi$ is the efficiency for detecting the products of decay $\chi$ for the reactions in which the outgoing ejectile $b$ is detected. The excitation energy $E^*$ is obtained by measuring the kinetic energies of the projectile beam and of the ejectile $b$, and the angle $\theta_{lab}$ between them in the laboratory reference frame.

Transfer and inelastic scattering reactions have been used since many years to measure fission probabilities $P_f(E^*)$ at the fission threshold and infer fission-barrier parameters, see e.g. [3, 4]. These fission probabilities have also often been employed to infer neutron-induced fission cross sections through the surrogate-reaction method [5]. In [6], we measured for the first time simultaneously the probabilities for γ emission, $P_\gamma$, and fission, $P_f$, of $^{240}$Pu, in an experiment where a $^{240}$Pu target was excited via the inelastic scattering of α particles. The measured probabilities allowed us to significantly reduce the uncertainty in some fission barrier parameters and to constrain the models for other nuclear properties like the GSF and NLD. Ultimately, we used the surrogate reaction method to determine the neutron-induced fission and radiative capture cross sections of $^{239}$Pu.

The measurement of de-excitation probabilities $P_\chi(E^*)$ in experiments in direct kinematics has important limitations: (i) when the nuclei of interest are far from stability, the necessary targets are unavailable. (ii) Competing reactions in target contaminants and backings produce a high background that is very complicated or even impossible to remove. (iii) The heavy products of the decay of the recoil nucleus $Y^*$ are stopped in the target sample and cannot be detected. Therefore, the measurement of neutron and γ-emission probabilities requires detecting the emitted neutrons and γ-rays, which is very complicated. To our knowledge, the neutron emission probability $P_n(E^*)$ has never been measured so far. In the case of $P_\gamma(E^*)$, the γ-ray-cascade detection efficiencies achievable are limited to a maximum of about 20 %, which leads to large statistical uncertainties [7].

Limitations (i) and (iii) can be addressed by conducting measurements in inverse kinematics with the heavy nucleus $X$ interacting as a projectile with the light nucleus $a$ at rest. Radioactive beam facilities can provide heavy-ion beams of very short-lived nuclei, and the heavy residues produced after γ and neutron emission can exit the target and are detectable. Nonetheless, the limited intensity of radioactive beams necessitates the use of thick targets. As a result, significant energy losses as well as energy and angular straggling in the target prevent the accurate determination of the beam and target residue energies, as well as the angle $\theta_{lab}$ of the target residue relative to the beam axis at the point of interaction. Accurate determination of these quantities is critical to determine the excitation energy of the excited nucleus $Y^*$ with sufficient resolution (a few hundred keV FWHM) to study the rapid evolution of the de-excitation probabilities at the particle or fission thresholds, see [6]. In addition, target windows and contaminants have to be avoided, as they can generate a strong background. This was the case in a recent experiment, where the fission probability of $^{239}$U was measured in inverse kinematics with the $^{238}$U(d,p) reaction using a $CD_2$ target. The carbon content of the $CD_2$ target led to a significant background in both the singles and fission coincidence spectra [8].

We aim to address the aforementioned target issues by measuring the decay probabilities $P_\chi(E^*)$ in inverse kinematics at a storage ring, for the first time. Our innovative approach, along with some of the results from our first experiment, are outlined in [9]. In this paper we provide a more comprehensive insight into our new experimental method and present some additional results. In Ref. [9], the neutron-induced radiative capture cross-section of $^{207}$Pb, $^{207}$Pb(n, γ), is inferred using our decay probabilities.

This work is structured as follows: in section II, we describe the advantages and challenges of heavy ion storage rings. In section III, we present our first experiment, in section IV we describe some particularities of two-body reactions in inverse kinematics and in section V we explain the data and uncertainty analysis. The results are discussed and compared with model calculations in section VI. The conclusion and outlook are given in section VII.

## II. Advantages and challenges of heavy ion storage rings

Heavy-ion storage rings are a type of circular lattice made up of bending and focusing multipolar magnetic elements (dipoles, quadrupoles, etc.) whose purpose is to store ions [10]. The electron cooler, which significantly reduces the size, angular divergence and momentum spread of the stored beam, is a key element of storage rings. If a gas-jet target is present inside the ring, the electron cooler can compensate for the energy loss and reduce the momentum and angular spread caused by the interaction of the beam with the target. In this way, the ion beam always reaches the target with the same energy and the same outstanding quality, and thus the energy loss and straggling effects that prevent the required $E^*$ resolution from being achieved can be neglected. In addition, the frequent passage of the target zone (around 1 million times per second at 10 MeV/nucleon) makes it possible to use pure gas-jet targets with ultra-low areal density (≈$10^{13}$ atoms/cm$^2$), and no windows are required.

Storage rings can also be used to reduce the energy of the stored beam from around 100 MeV/nucleon, which is the typical energy required to produce bare ions, to a few MeV/nucleon. This enables another unique feature: the preparation and storage of cooled beams of fully stripped radioactive heavy ions [11] at about 10 MeV/nucleon. This feature is particularly advantageous for the study of transfer or inelastic scattering reactions compared to single-pass experiments at fragmentation facilities with fast radioactive beams of about 100 MeV/nucleon, where the beam energy resolution and the emittance are degraded, the cross-sections are reduced [12] and the kinematic compression [13] is increased.

Moreover, since the gaseous target has a very low density, the likelihood of electron capture reactions prior to or following the nuclear reaction is extremely low. As a result, the beam-like residues produced by the nuclear reaction will also be completely stripped. This is beneficial for our measurements as it allows for the detection of beam-like residues with a very large efficiency, as explained in section V. All these advantages are only possible with storage rings, which enable the determination of the excitation energy and decay probabilities of the decaying nucleus with unparalleled precision.

However, to avoid beam intensity losses due to atomic interactions of the stored ion beam with the residual gas in the ring, heavy ion storage rings must be operated under ultra-high vacuum (UHV) conditions ($10^{-10}$ to $10^{-12}$ mbar), which places severe constraints on the detection systems located in the ring. Only recently, UHV-compatible silicon detectors have been used in pioneering experiments for the study of nuclear reactions [14, 15, 16] at the experimental storage ring (ESR) [17] and at the CRYRING storage ring [18] of the GSI/FAIR facility.

## III.    Experiment at the ESR

We performed our experiment at the ESR, where we used the inelastic scattering reaction $^{208}$Pb(p,p') to excite a $^{208}$Pb$^{82+}$ beam and measure its γ and neutron-emission probabilities. The elastic scattering reaction $^{208}$Pb(p,p) was also particularly interesting for us because the ground state of $^{208}$Pb is separated by 2.61 MeV from its first excited state. Therefore, the width of the elastic scattering peak can provide direct information on the excitation energy resolution $\Delta E^*$, which is one of the important aspects we wanted to investigate in this first experiment.

The beam preparation was as follows: a pulse of about $10^8$ bare $^{208}$Pb$^{82+}$ ions was injected into the ESR at 271 MeV/nucleon. The ions were cooled and decelerated to 30.77 MeV/nucleon in about 25 s, then the hydrogen gas-jet target was turned on and the measurement started for about 30 s. After the measurement period, the target was turned off, the beam was dumped away, and a new pulse was injected. On average, $5 \cdot 10^7$ $^{208}$Pb$^{82+}$ ions were cooled and decelerated per injection, and revolved with a frequency of 0.695 MHz. The average target areal density was $6 \cdot 10^{13}$ atoms/cm$^2$.

To determine the $P_\gamma(E^*)$ and $P_n(E^*)$ probabilities of $^{208}$Pb*, we measured the scattered protons with a ΔE-E Si telescope and the beam-like residues produced after the de-excitation of $^{208}$Pb* via γ-ray and neutron emission with a position-sensitive Si strip detector (denoted beam-like residue detector in Fig. 1) placed behind the ring dipole magnet downstream of the target. This dipole magnet acted as a recoil spectrometer separating the unreacted beam, the $^{208}$Pb$^{82+}$ residues produced after γ-ray emission and the $^{207}$Pb$^{82+}$ residues produced after neutron emission, see Fig. 1. The high rate of $^{208}$Pb$^{81+}$ residues produced after electron capture of the beam in the target was not problematic because these residues possessed a smaller charge, resulting in a larger magnetic rigidity, and were bent outside the ring, away from the $^{208}$Pb$^{82+}$ and $^{207}$Pb$^{82+}$ residues.

To prevent detector components from degrading the UHV of the ring, the telescope and the beam-like residue detector were housed in pockets with 25-µm thick stainless steel windows through which the scattered protons and beam-like residues could pass. Both pockets were operated in air at atmospheric pressure. The telescope was centred at 60° with respect to the beam axis at a distance of 101.3 mm from the target. This distance was chosen to avoid beam interception by the telescope after injection, where significant fluctuations in the horizontal beam position are possible. The ΔE telescope detector consisted of a 530 µm thick 20×20 mm$^2$ double-sided silicon strip detector (DSSD) with 16 vertical and 16 horizontal strips at a pitch of 1250 µm, enabling energy loss and $\theta_{lab}$ to be measured. The ΔE detector was followed by the E detector, consisting of a stack of six single-area detectors for total energy measurements. Each of the detectors of the stack had an active area of 20×20 mm$^2$ and a thickness of 1.51 mm. The total thickness of the telescope, 9.6 mm, was sufficient to stop scattered protons up to 43 MeV. The angular coverage of the telescope in the laboratory reference frame ranged from $\theta_{lab}$=54.8 to 64.6°.

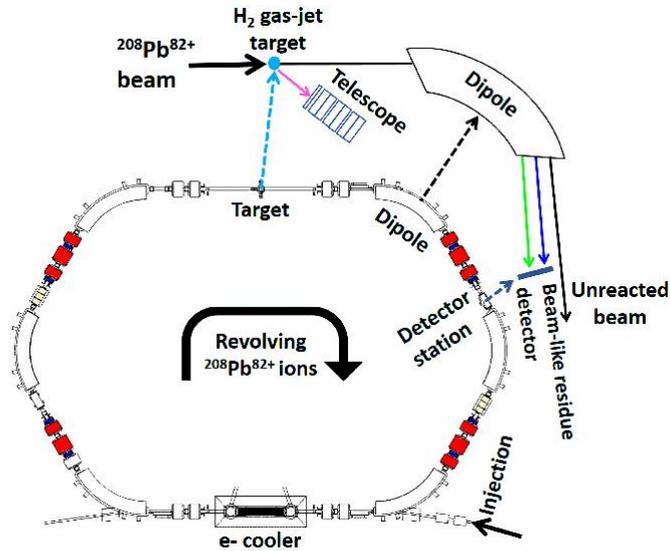

**Figure 1:** The lower part shows a schematic view of the ESR. The upper part shows the portion of the ring where our detectors were installed. The trajectory of the scattered protons is represented by the pink arrow and the one of the beam by the black arrow. The trajectories of the heavy beam-like residues produced after emission of γ rays ($^{208}$Pb$^{82+}$) and a neutron ($^{207}$Pb$^{82+}$) are indicated by the blue and green arrows, respectively.

The beam-like residue detector was a DSSD with a thickness of 500 µm, an active area of 122×40 mm², 122 vertical strips and 40 horizontal strips. The pocket housing this detector was installed in a movable drive, which moved the pocket in when the target was switched on after beam cooling and deceleration, and moved it out at the end of the measurement cycle. In this way, we protected the detector from accidental interactions with the uncooled beam. When moved in, the detector was positioned 15.0 ± 0.1 mm from the beam axis. This distance ensured that the rate of elastically-scattered beam ions over the entire detector was well within the tolerance range for radiation damage of the detector, which remained operational for the duration of the experiment.

The detector signals were readout with the commercially available Multi-channel Multiplexed Read-out system (MMR) by Mesytec [19]. Our system was composed of several front-end modules MMR located nearby the detectors, which were connected to central VME data collectors (the VMMR) through optical fibers. At the front-end, the detector signals were preprocessed during a user-defined time window and the first channel hit provided a trigger request out of the VMRR. The trigger request was sent to a trigger logic module, which took the read-out decision. If the trigger was accepted, the data were formatted into an event structure and sent to a buffer for storage.

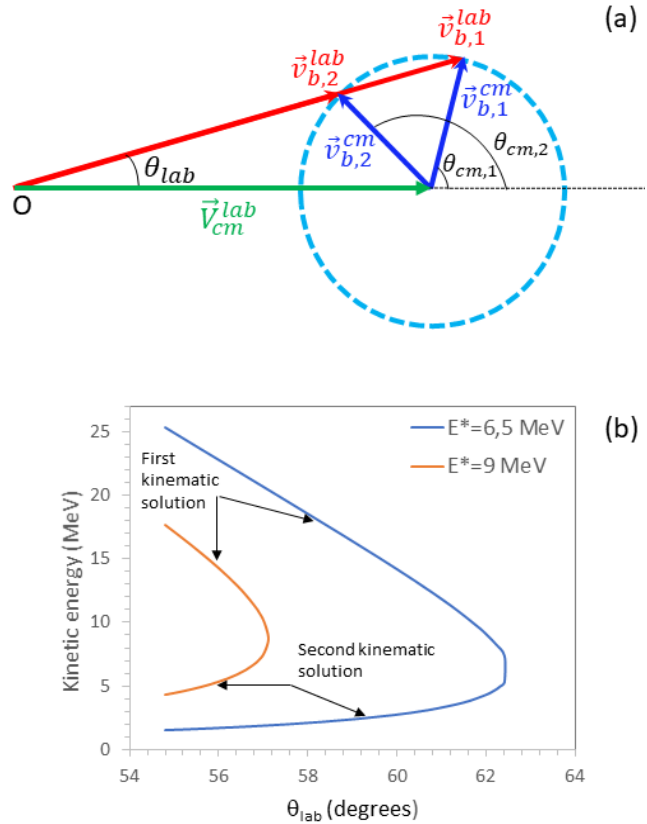

**Figure 2:** (a) Non relativistic velocity addition diagram for an inelastic scattering reaction in inverse kinematics. The radius of the circle represents the length of the velocity vector of the ejectiles $b$ in the center of mass for a given excitation energy. See text for details. (b) Kinetic energy of the scattered protons as a function of their emission angle in the laboratory $\theta_{lab}$ for the $^{208}$Pb(p,p') reaction at 30.77 MeV/nucleon and two excitation energies.

## IV. Two-body reactions in inverse kinematics

In inverse kinematics there is not always a one-to-one relation between the kinetic energy of the ejectile $b$ and the angle $\theta_{lab}$. It is possible to have two groups of ejectiles $b$ with two different kinetic energies and the same $\theta_{lab}$. These two groups are related to two kinematic solutions and their origin can be most easily understood by considering a generic, non-relativistic velocity diagram like the one shown on panel (a) of Fig. 2. The velocity of the ejectiles in the laboratory $\vec{v}_b^{lab}$ is given by the sum of the velocity of the ejectiles in the center of mass reference system $\vec{v}_b^{cm}$ and the velocity of the center of mass in the laboratory frame $\vec{V}_{cm}^{lab}$. We can see that at the angle $\theta_{lab}$ two velocities of the ejectile $b$ in the laboratory are possible that result from the emission of the ejectile $b$ at two different center of mass angles $\theta_{cm}$. In panel (b) of Fig. 2, the kinetic energies in the laboratory of scattered protons from the $^{208}$Pb(p, p') reaction are shown as a function of $\theta_{lab}$ for two different excitation energies. We can see that for a given $\theta_{lab}$ the kinetic energy of scattered protons from the second kinematic solution is much lower than for the protons from the first kinematic solution.

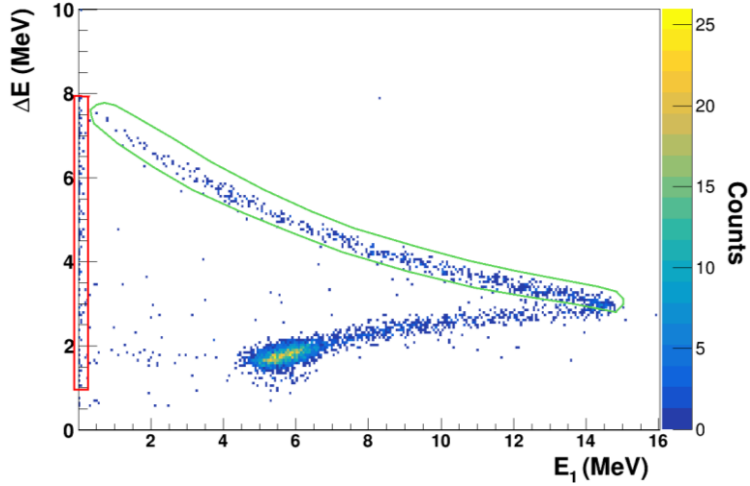

**Figure 3:** Energy loss in the vertical strip centered at $\theta_{lab}$ = 58.9° of the ΔE detector of the telescope versus the energy deposited in first detector (E$_1$) of the E detector stack by scattered protons from the $^{208}$Pb(p,p') reaction. The protons from the second kinematic solution are stopped in the ΔE detector and are within the red contour. The protons from the first kinematic solution that are stopped in the E$_1$ detector of the E-detector stack are inside the green contour.

## V. Data analysis

Fig. 3 shows the energy loss in a strip of the ΔE detector centered at $\theta_{lab}$= 58.9° versus the energy deposited in the first detector of the E-detector stack, the E$_1$ detector. The protons from the second kinematic solution are stopped in the ΔE detector and are inside the red contour. They are emitted within the angular range in the center of mass $\theta_{cm}$=152-167°. The results for the neutron emission probability obtained with these events are presented in [9]. The protons from the first kinematic solution, which are stopped in the E$_1$ detector are located within the green contour. In this work, we will concentrate on these events, which correspond to inelastic scattered events in the center of mass angular range $\theta_{cm}$=122-147° and the $E^*$ range between $E^*$=6.5 and 9.3 MeV. This is the $E^*$ range of interest in this work, where the γ and neutron-emission de-excitation modes of $^{208}$Pb* compete. The protons outside the green and red contours have sufficient energy to punch through the E$_1$ detector. The most energetic protons are the elastic scattered protons, which lead to a clearly visible peak at ΔE ≈ 1.8 and E$_1$ ≈ 5.8 MeV.

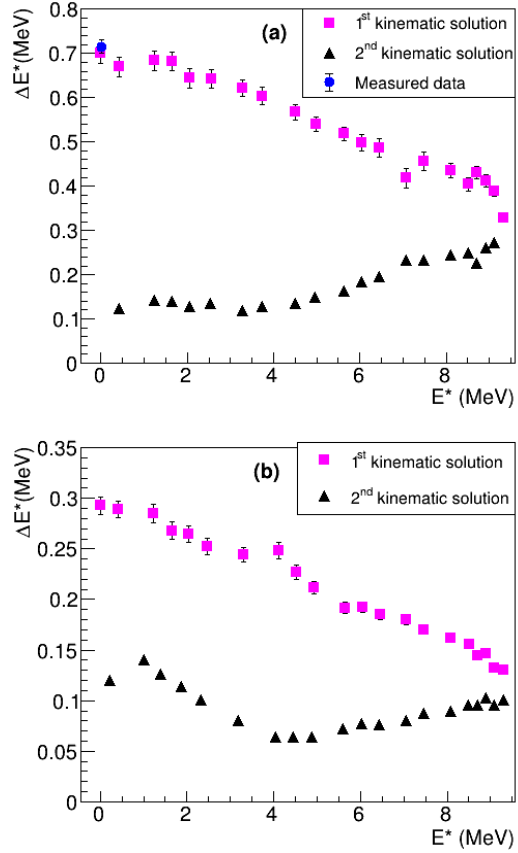

**Figure 4:** Results of our simulations for the excitation energy resolution $\Delta E^*$ (RMS) as a function of the excitation energy $E^*$ for the first and second kinematic solutions of the $^{208}$Pb(p,p') reaction. (a) Results for a target radius of 2.5 mm. The blue triangle at $E^* = 0$ MeV represents the experimental result for $\theta_{lab}$=61°. (b) Results for a target radius of 0.5 mm.

### A. Excitation energy resolution

We calibrated the detectors of the telescope in energy with the elastic scattered peaks measured at three different beam energies 30.77, 36.91 and 43.04 MeV/nucleon. Using the energy deposited in the telescope by the elastically scattered protons at 30.77 MeV/nucleon and the detection angle $\theta_{lab}$ of the corresponding strip with respect to the center of the target, we could infer the excitation energy. The root-mean squared (RMS) deviation of the elastic peak for the strip centered at 61° amounts to $\Delta E^* = 726 \pm 15$ keV. We developed a Geant4 [20] simulation that considers all the effects that impact $\Delta E^*$ such as the target radius of 2.5 mm, the segmentation of the ΔE detector, the energy resolution of the telescope detectors and the momentum spread and emittance of the beam. Our simulation gives $\Delta E^* = 698 \pm 22$ keV at $E^*$=0 MeV, which agrees very well with the measured value. This agreement validates our simulations, which we have then used to infer $\Delta E^*$ at higher $E^*$. Panel (a) of Fig. 4 shows that for the first kinematic solution $\Delta E^*$ decreases with increasing $E^*$, and varies from 420 ± 22 to 330 ± 8 keV (RMS) between $E^*$=7 and $E^*$=9.3 MeV. For the second kinematic solution the average value of $\Delta E^*$ is about 240 keV (RMS) for $E^*$ from 6.5 to 9.1 MeV. The uncertainties in the simulated values include only statistical fluctuations. Panel (b) of Fig. 4 shows that $\Delta E^*$ can be significantly improved with a smaller target radius of 0.5 mm, reaching values of $\Delta E^* \approx$ 130 keV for the first kinematic solution and below 100 keV for the second kinematic solution. This reflects that in the present experiment $\Delta E^*$ is dominated by the target radius, which induces a significant uncertainty on the emission angle $\theta_{lab}$. The resolution in $E^*$ is better for the second kinematic solution because this kinematic solution shows a weaker dependence of the kinetic energy of the scattered protons on the emission angle $\theta_{lab}$, as can be seen on Fig. 2 (b).

### B. Determination of single and coincidence events

The singles spectrum represents the number of detected protons as a function of the $E^*$ of $^{208}$Pb. It is obtained by selecting the protons inside the green contour of Fig. 3 and by using the telescope strip angles $\theta_{lab}$ and the measured proton kinetic energy to infer the $E^*$ of $^{208}$Pb event by event. The singles spectrum $N_s(E^*)$ for the strip centered at $\theta_{lab} = 58.9°$ is shown in black in Fig. 5. The bin size of this histogram is 200 keV.

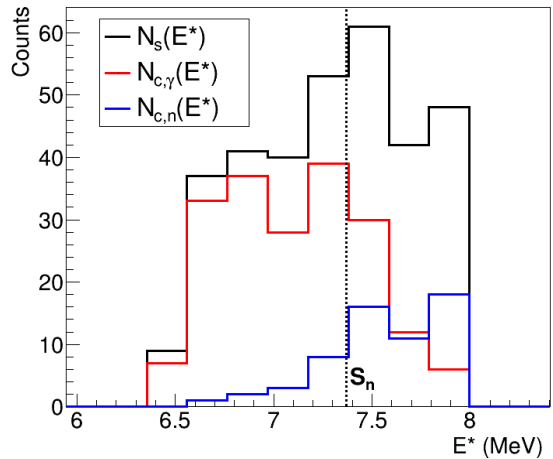

**Figure 5:** Singles ($N_s(E^*)$, black) and coincidence spectra for γ ($N_{c,\gamma}(E^*)$, red) and neutron ($N_{c,n}(E^*)$, blue) emission measured for the $^{208}$Pb(p,p') reaction at $\theta_{lab} = 58.9°$ and scattered protons inside the green contour of Fig. 3. The vertical dotted line indicates the neutron separation energy $S_n$ of $^{208}$Pb, $S_n$=7.37 MeV.

Figure 6 (a) shows a scatter plot representing the position of the heavy ions impinging in the beam-like residue detector. This plot is dominated by the intense rate of elastically scattered beam ions in the target. In parts 6 (b) and 6 (c) we see the same plot but for events measured in coincidence with scattered protons detected in the telescope. It can be seen that the coincidence requirement drastically suppresses the background due to elastic scattering, resulting in the emergence of the position peaks of the beam-like reaction residues. On panel (b) we see the heavy residues measured in coincidence with protons from the first kinematic solution corresponding to $E^*$=5.6-9.5 MeV and $\theta_{lab}$=55.4-60.5°. We can clearly distinguish two peaks; the left peak contains the $^{208}$Pb$^{82+}$ nuclei formed after γ emission and the right peak the $^{207}$Pb$^{82+}$ nuclei produced after neutron emission. On panel (c) are shown the heavy residues detected in coincidence with protons from the second kinematic solution. In this case, the residues of $^{208}$Pb$^{82+}$ produced after γ emission have larger kinetic energies than those of the first kinematic solution, their trajectories after the dipole magnet are very close to the beam axis and cannot be detected. Therefore, we only see the peak containing the $^{207}$Pb$^{82+}$ residues.

The coincidence spectra $N_{c,\gamma}(E^*)$ and $N_{c,n}(E^*)$ are obtained in the same way as the singles spectra but for protons detected in coincidence with beam-like residues inside the red and the dashed blue contours of Fig. 6 (b), respectively. The two spectra are shown in red and blue in Fig. 5. After correction for the detection efficiency, these spectra represent, respectively, the number of $^{208}$Pb* nuclei that have decayed by γ and neutron emission for the reactions where the scattered protons were detected. We have used eq. (1) to infer the de-excitation probabilities at different $\theta_{lab}$, i.e. we calculated for each telescope vertical strip the ratio between the coincidence spectrum $N_{c,\chi}(E^*)$ and the product of the singles spectrum $N_s(E^*)$ and the detection efficiency of the beam-like residues $\varepsilon_\chi$.

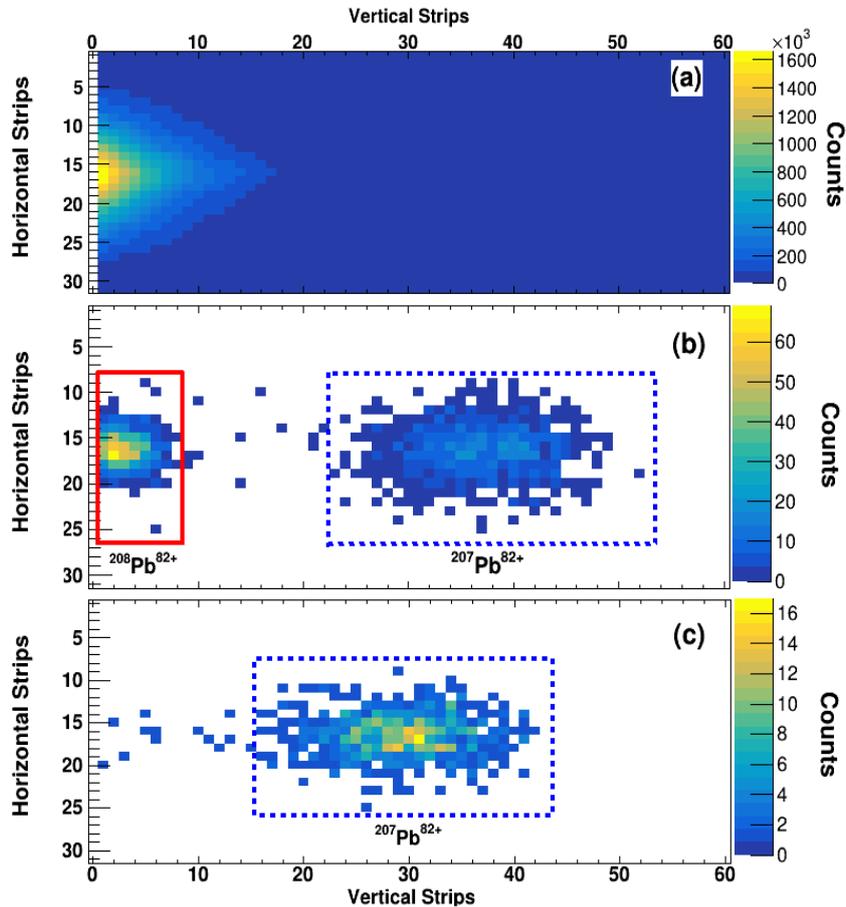

**Figure 6:** Position of the heavy ions detected by the beam-like residue detector without coincidence with the telescope (a), in coincidence with scattered protons from the first kinematic solution (b) and in coincidence with protons from the second kinematic solution (c). The red contour in panel (b) includes the events that de-excite by the emission of γ rays, while the blue dashed contours in panels (b) and (c) contain the events that de-excite by the emission of a neutron.

## C. Detection efficiency of the beam-like residues

The detection efficiency for the beam-like residues is determined by the transmission efficiency of the heavy residues between their production site in the gas-jet target and the position of the beam-like residue detector, as well as by the intrinsic and the geometric efficiencies of the beam-like residue detector.

The transmission efficiency was determined with our Geant4 simulation by computing the trajectories of the $^{208}$Pb$^{82+}$ and $^{207}$Pb$^{82+}$ residues produced in coincidence with scattered protons detected in the telescope. The simulation included the relevant portion of the ESR lattice, extending from the target to the beam-like residue detector. The results of the simulation demonstrated that there were no collisions between the heavy lead residues and beam pipes or electromagnetic elements of the ring, indicating that all of them could reach the beam-like residue detector plane. As previously stated, the probability of efficiency losses due to atomic reactions in the target before or after the nuclear reaction is considered to be extremely low. Our estimation of this probability is in the range of $10^{-20}$. Furthermore, the probability of losses from electron capture in the residual gas between the target and the heavy-residue detector is estimated to be as low as $10^{-14}$. The intrinsic detection efficiency of the beam-like residue detector for the $^{208}$Pb and $^{207}$Pb residues is 100%. This is due to the significant energy deposited by these residues, resulting in induced signal amplitudes easily exceeding the electronic threshold, even for interstrip events. Consequently, in this experiment, the detection efficiency of the beam-like residues is defined by the geometric efficiency of the beam-like residue detector, that is to say, by the extent to which the specific heavy-residue emission cone is covered by the active area of the detector.

Figure 6 (b) shows that part of the γ-emission peak is outside the beam-like residue detector, which implies that in some cases the geometric efficiency for γ emission is less than 100%. On the other hand, the shape of the neutron emission peak in Fig. 6 (b) indicates that all the trajectories of the $^{207}$Pb residues hit the beam-like residue detector, because the $^{207}$Pb residues are more deflected by the dipole magnet than the $^{208}$Pb residues due to their lower mass. This was confirmed by our simulations.

To validate the results of our simulations for the efficiency of the γ-emission channel, $\varepsilon_\gamma$, we used the data measured at $E^*$ below the neutron separation energy $S_n$ of $^{208}$Pb, $S_n$=7.37 MeV. At these excitation energies, $\varepsilon_\gamma$ can be deduced from the ratio of coincidence over single events because the γ-ray emission probability $P_\gamma(E^*)$ is equal to 1 and from eq. (1) it results that $\varepsilon_\gamma(E^*)=N_{c,\gamma}(E^*)/N_s(E^*)$. The simulated and experimental values for $\varepsilon_\gamma$ obtained for one telescope strip are compared in Fig. 7, showing a good agreement. In this figure, the efficiency decreases with increasing $E^*$ because the kinetic energy of the $^{208}$Pb residues increases. The residues are therefore less deflected by the dipole magnet, which leads to larger detection losses since their trajectories are closer to the beam axis, i.e. farther away from the beam-like residue detector.

The absolute uncertainty in $\varepsilon_\gamma$ varies from $\Delta\varepsilon_\gamma$ = 2 to 6%. It is given by fluctuations of the horizontal beam position on the target, which led to fluctuations in the position of the beam-like residues at the detection plane. We observed that the average position of the heavy residues fluctuated between two adjacent detector strips over the measurement time. These fluctuations were propagated into the uncertainty in $\varepsilon_\gamma$ by changing the horizontal position of the beam-like residue detector in the simulation by ± 0.35 mm. In this experiment $\varepsilon_\gamma$ varied between 33 and 100% depending on $E^*$ and $\theta_{lab}$.

The detection efficiency for the neutron emission channel, $\varepsilon_n$, is 100%, regardless of the scattering angle $\theta_{lab}$ and the excitation energy $E^*$. The fluctuations in the beam position at the target do not affect $\varepsilon_n$. Therefore, the uncertainty $\Delta\varepsilon_n$ can be neglected, as it is only given by fluctuations in the losses of beam-like residues due to atomic reactions in the target and/or the residual gas. The fact that $\varepsilon_n = 100\%$ and $\Delta\varepsilon_n \approx 0$ demonstrates the considerable advantages of our novel methodology compared to measurements in direct kinematics and to single-pass experiments in inverse kinematics. Single-pass experiments require thick targets, which results in the formation of projectile-like residues with varying charge states and significantly lower detection efficiencies.

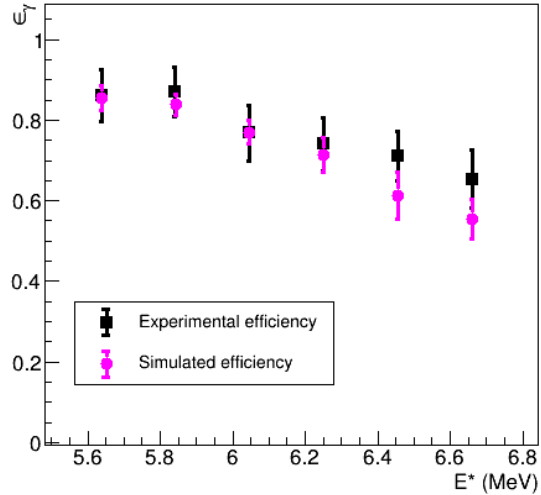

**Figure 7:** Detection efficiency for the γ-emission channel $\varepsilon_\gamma$ as a function of the excitation energy of $^{208}$Pb for $\theta_{lab}=60.4°$. The black squares represent the experimental results and the pink circles the simulated results.

### D. Uncertainty analysis

Applying error propagation to eq. (1) we obtain:

$$\left(\frac{\Delta P_\chi(E^*)}{P_\chi(E^*)}\right)^2 = \left(\frac{\Delta N_{c,\chi}(E^*)}{N_{c,\chi}(E^*)}\right)^2 + \left(\frac{\Delta N_s(E^*)}{N_s(E^*)}\right)^2 + \left(\frac{\Delta \varepsilon_\chi(E^*)}{\varepsilon_\chi(E^*)}\right)^2 - 2 \cdot \frac{Cov(N_{c,\chi}(E^*), N_s(E^*))}{N_{c,\chi}(E^*) \cdot N_s(E^*)} \quad (2)$$

where $\Delta N_{c,\chi}$ and $\Delta N_s$ are the uncertainties in the number of coincidence and single events and $Cov(N_{c,\chi}, N_s)$ is the covariance term between $N_{c,\chi}$ and $N_s$. In equation (2) we have omitted the covariance terms involving the efficiency $\varepsilon_\chi$ because the efficiencies have been determined with our simulation and are independent from all the measured quantities.

As already highlighted in our previous work [4, 21], eq. (2) shows that the covariance term between $N_{c,\chi}$ and $N_s$ can significantly reduce the uncertainty in $P_\chi$. The covariance term $Cov(N_{c,\chi}, N_s)$ measures how fluctuations in $N_s$ affect the value of $N_{c,\chi}$. In [4] we deduced analytically that:

$$Cov(N_{c,\chi}, N_s) = (\Delta N_{c,\chi})^2 = N_{c,\chi} \qquad (3)$$

where we have considered that $\Delta N_{c,\chi} = \sqrt{N_{c,\chi}}$. To derive eq. (3) some approximations were used based on the detection efficiencies of the direct-kinematics experiment described in [4]. To verify that eq. (3) is also valid in the present case, we have inferred $Cov(N_{c,\chi}, N_s)$ by using an alternative procedure, which we introduced in [21]. In this procedure, we consider groups of measured single events containing a number of events $N_s^i$ that fluctuates following a Gaussian distribution with mean value $\omega$ and standard deviation $\sqrt{\omega}$. The number of measured coincidence events associated to each group $i$ of single events, $N_{c,\chi}^i$, is then represented versus the number of single events $N_s^i$, see Fig. 8. In this figure, we have considered a fluctuating number of single events in the excitation energy range from 7.6 MeV to 7.8 MeV with $\omega = 34$. Using the data of Fig. 8, we have determined the value of $Cov(N_{c,\chi}, N_s)$ by applying the equation:

$$Cov(N_{c,\chi}, N_s) = \frac{1}{\mathbb{N}} \sum_{i=1}^{\mathbb{N}} (N_{c,\chi}^i - \langle N_{c,\chi} \rangle) \cdot (N_s^i - \langle N_s \rangle) \qquad (4)$$

where $\langle N_{c,\chi} \rangle$ and $\langle N_s \rangle$ are the mean values of $N_{c,\chi}^i$ and $N_s^i$, respectively, and $\mathbb{N}$ is the number of groups of data sampled, in our case $\mathbb{N} = 20000$. For determining $(\Delta N_{c,\chi})^2$ we used:

$$(\Delta N_{c,\chi})^2 = \frac{1}{\mathbb{N}} \sum_{i=1}^{\mathbb{N}} (N_{c,\chi}^i - \langle N_{c,\chi} \rangle)^2 \qquad (5)$$

We obtained $Cov(N_{c,\gamma}, N_s) = 10.25$ and $Cov(N_{c,n}, N_s) = 10.92$, which are comparable with $(\Delta N_{c,\gamma})^2 = 9.98$ and $(\Delta N_{c,n})^2 = 10.71$, respectively, thus validating equation (3). The correlation between $N_{c,\chi}$ and $N_s$ is defined as $\rho_\chi = \frac{Cov(N_{c,\chi}, N_s)}{\sqrt{(\Delta N_{c,\chi})^2 \cdot (\Delta N_s)^2}}$ and has a maximum value of 1. In this experiment, it amounts to $\rho_\gamma = 0.62$ and $\rho_n = 0.64$, and can be clearly seen in Fig. 8.

Inserting eq. (3) in eq. (2) and using $(\Delta N_{c,\chi})^2 = N_{c,\chi}$ and $(\Delta N_s)^2 = N_s$ we get:

$$\Delta P_\chi(E^*) = \sqrt{\left(P_\chi(E^*)\right)^2 \cdot \left(\frac{1}{N_{c,\chi}(E^*)} + \frac{1}{N_s(E^*)} + \left(\frac{\Delta \varepsilon_\chi(E^*)}{\varepsilon_\chi(E^*)}\right)^2 - 2 \cdot \frac{1}{N_s(E^*)}\right)} \qquad (6)$$

The number of coincidence events equals the number of single events when the decay probability is 1 and the detection efficiency is 100 %. In this case, it follows from eq. (6) that the uncertainty in the decay probability is solely determined by the uncertainty in the detection efficiency. Our experiment is characterized by very large and precise efficiencies (particularly for the neutron-emission channel where $\varepsilon_n = 100\%$ and $\Delta \varepsilon_n \approx 0$). This will lead to rather low uncertainties when the probabilities are high, even with limited statistics.

We have added quadratically to the numerator of the first term in eq. (2) the quantity $\Delta N_{c,\chi}^r$, which corresponds to the uncertainty in the random coincidences with the elastically scattered beam, $\Delta N_{c,\chi}^r = \sqrt{N_{c,\chi}^r}$. The random coincidences $N_{c,\chi}^r$ can be seen on the left side of the blue contour in Fig. 6 (c) and between the red and blue contours in Fig. 6 (b). The number of these events is rather small, corresponding to 17 and 14 counts, respectively, for the full statistics of the experiment. To determine the number of random events in the $^{208}$Pb and $^{207}$Pb position peaks, we used the random coincidence events in Fig. 6 (c). In the case of the $^{207}$Pb peak, we applied a reduction factor derived from the position spectrum of the heavy residues measured without coincidence with the scattered protons shown in Fig. 6 (a). The profile of this spectrum exhibits a typical elastic scattering pattern, characterized by a pronounced intensity drop as the distance to the detector edge increases. The number of random coincidences per telescope strip and bin in $E^*$, $N_{c,\chi}^r$, represents at most a few counts and has an impact on the uncertainty of $P_\chi(E^*)$ only when $N_{c,\chi}$ tends to zero, i.e. at the highest $E^*$ for $P_\gamma$ and near the neutron threshold $S_n$ for $P_n$.

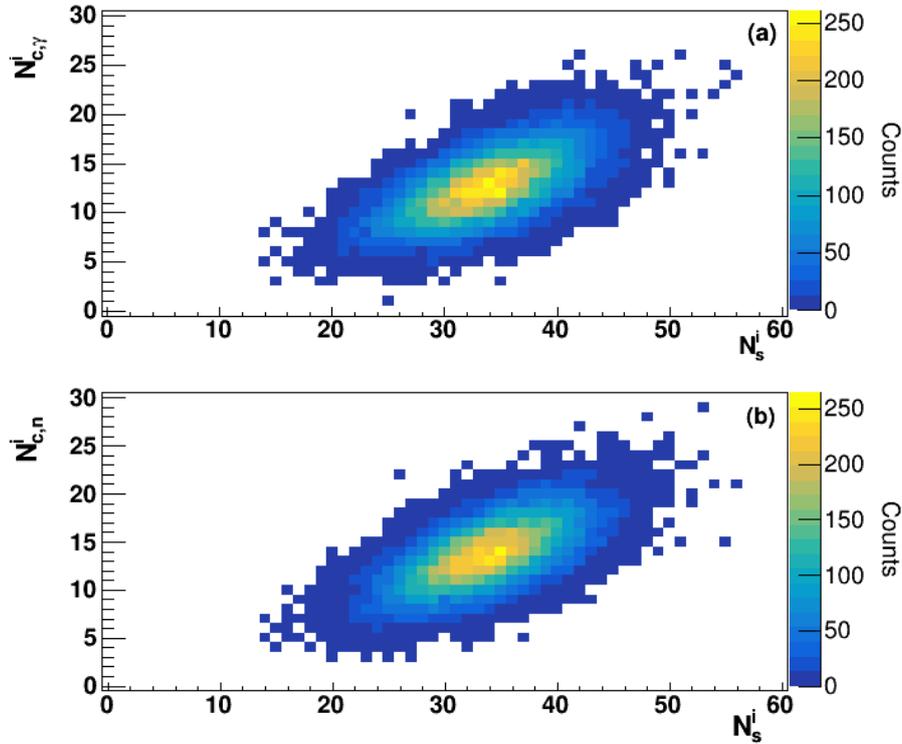

**Figure 8:** Number of measured coincidence events $N_{c,\gamma}^i$ (panel a) and $N_{c,n}^i$ (panel b) versus the number of single events $N_s^i$ in the excitation energy range from 7.6 MeV to 7.8 MeV. The number of single events has been sampled 20000 times from a Gaussian distribution with mean value 34 and standard deviation $\sqrt{34}$, see text for details.

## VI. Results

The probabilities obtained with the individual telescope strips agree within the error bars. Therefore, the final probabilities were determined by calculating the weighted mean of the probabilities per strip. Figure 9 shows the resulting $\gamma$- and neutron-emission probabilities as a function of $E^*$. It can be seen that $P_\gamma$ is 1 at the lowest $E^*$ and starts to decrease near $S_n$, due to the competition with neutron emission, whose probability $P_n$ begins to differ from zero. Since γ and neutron emission are the only open de-excitation channels within the covered excitation energy range, the sum of the two probabilities has to be equal to 1. This condition is well satisfied by our data as reflected by the black dots. This result validates our new methodology and in particular the determination of the efficiencies $\varepsilon_\chi$. The $E^*$ at which $P_n$ starts to increase is below $S_n$. As we will see later, this is due to the excitation energy resolution. Between $S_n$ and $E^* = 9.1$ MeV the relative uncertainty of $P_\gamma(E^*)$ increases from 9 to 40 %, while for $P_n(E^*)$ the relative uncertainty decreases from 14 to less than 3 %. The uncertainties for $P_\gamma$ are larger than for $P_n$ because of the larger uncertainty in the detection efficiency $\varepsilon_\gamma$ of the γ-emission channel.

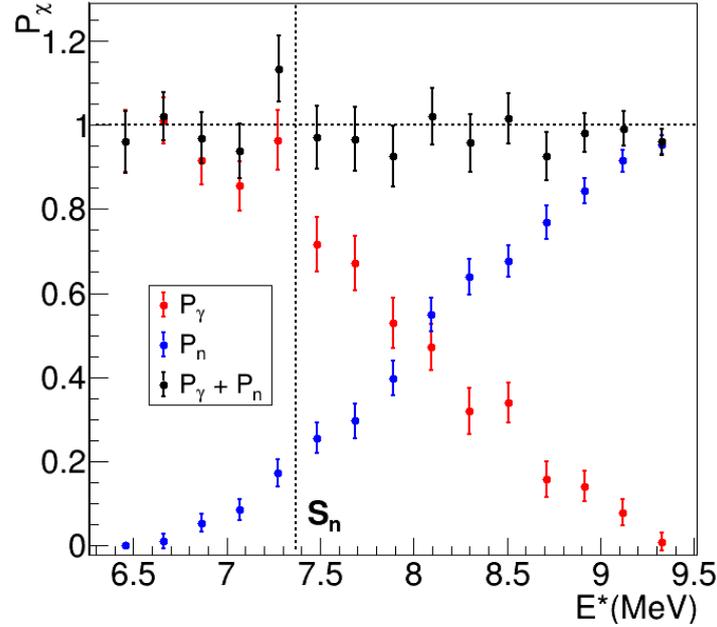

**Figure 9**: Probabilities for γ (red circles) and neutron (blue circles) emission as a function of the excitation energy $E^*$ of $^{208}$Pb. The black points represent the sum of the γ- and neutron-emission probabilities. The vertical line indicates the neutron separation energy $S_n$ of $^{208}$Pb, $S_n$=7.37 MeV, and the constant horizontal line at $P_\chi = 1$ is to guide the eye.

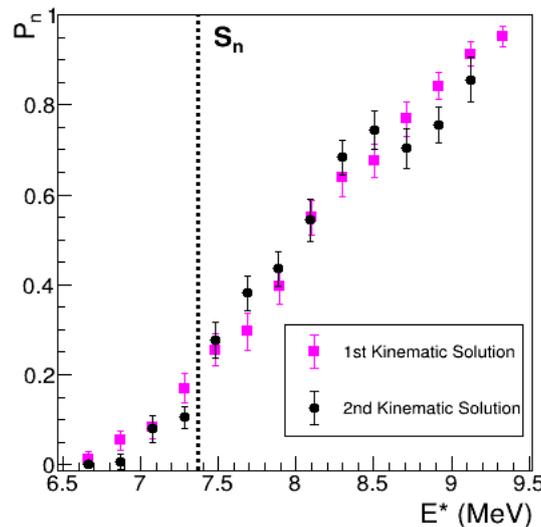

**Figure 10:** Neutron emission probabilities as a function of the excitation energy of $^{208}$Pb obtained with the first (pink squares) and the second (black circles) kinematic solutions. The vertical line indicates the neutron separation energy $S_n$ of $^{208}$Pb, $S_n$=7.37 MeV.

A comparison of the neutron emission probability obtained in this work with the result obtained with the second kinematic solution [9] is presented in Fig. 10. Overall, there is a good agreement between the two results, although there are some differences, for example, near $S_n$. These differences can be due to statistical fluctuations, to the $E^*$ resolution, which we have seen is worse for the first kinematic solution, and to the populated angular momentum and parity distributions, which can diverge because they depend on the proton scattering angle in the center of mass, $\theta_{cm}$. The calculations shown in Fig. 14 in the next section provide a potential explanation for these differences.

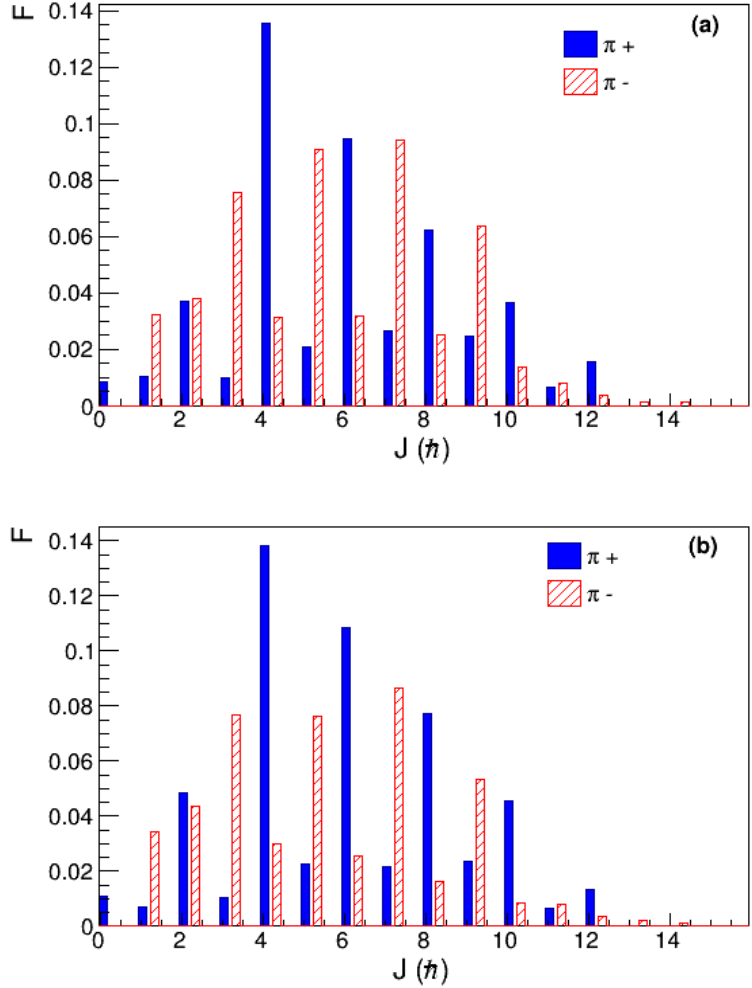

**Figure 11:** Calculated angular momentum $J$ and parity $\pi$ distributions populated in the $^{208}$Pb(p,p') reaction at 30.77 MeV/nucleon for $E^* = 8$ MeV. In panel (a) are the results for the first kinematic solution (a) and in panel (b) for the second kinematic solution.

### A. Comparison with model calculations

Within the frame of the statistical model, the decay probabilities are given by the expression:

$$P_\chi(E^*) = \sum_{J^\pi} F(E^*, J^\pi) \cdot G_\chi(E^*, J^\pi) \tag{7}$$

where $F$ is the probability that the used reaction leads to the formation of a compound nucleus in a state with angular momentum $J$ and parity $\pi$ at excitation energy $E^*$. $G_\chi$ is the probability that the compound nucleus decays from that state via de-excitation channel $\chi$. As detailed in [9], the $J^\pi$ distributions $F$ were calculated with the formalism described in refs. [22, 23]. We have calculated the $F(E^*, J^\pi)$ distributions from $E^*= 1$ to 9.5 MeV in steps of 0.5 MeV. The $F(E^*, J^\pi)$ distributions vary smoothly with $E^*$, and this variation has been considered in the calculation of $P_\chi(E^*)$ by implementing the F($E^*, J^\pi$) distributions at the different $E^*$ in eq. (7). Our calculations show that the $J^\pi$ distributions for the two kinematic solutions are very similar. For example, at $E^*= 8$ MeV the average angular momentum for positive parities amounts to 5.8 ℏ for the first and for the second kinematic solutions. For negative parities the average spins are 5.6 and 5.3 ℏ for the first and second kinematic solution, respectively. The calculated $J^\pi$ distributions at $E^* = 8$ MeV are shown in Fig. 11.

**Table 1:** Models used for the level-density and γ strength function of $^{208}$Pb, see text for details.

| Nuclear level density | γ-ray strength function |
|---|---|
| CT1, $T$=0.92 MeV, $E_0$=1.37 MeV from [24] | KU with parameters from [26] |
| CT2, $T$=0.82 MeV, $E_0$=1.81 MeV from [26] | SMLO with parameters from [26] |
| CT3, $T$=0.69 MeV, $E_0$=1.67 MeV from [25] | Microscopic D1M+QRPA [34] |
| BSFG [28] | |
| Microscopic Goriely et al. [30] | |
| Microscopic Hilaire et al. [31] | |

The probabilities $G_\chi$ were obtained with the Hauser-Feshbach statistical model implemented in TALYS 1.96 [26]. The two most uncertain ingredients for determining the probabilities $G_\chi$ are the nuclear level densities (NLD) and the γ-ray strength functions (GSF). As in [9], we have used several models for these two quantities with parameters for $^{208}$Pb found in the literature. The different descriptions are listed in Table 1. Three descriptions of the NLD are based on the constant temperature (CT) model [27], which has two parameters $T$ and $E_0$. These descriptions are denoted CT1, CT2 and CT3. The values of $T$ and $E_0$ for each description and the references from which they were taken are given in Table 1. We also considered the experimental NLD of $^{208}$Pb measured by Bassauer et al. [28], which is described by the back-shifted Fermi gas (BSFG) model [29]. The other two NLD descriptions are based on microscopic calculations by Goriely et al. [30] and by Hilaire et al. [31], the results of which are given in tabular form. Goriely et al. [30] employ the effective Skyrme interaction BSk14, while Hilaire et al. [31] utilize the D1M Gogny interaction. The BSFG and the CT3 NLDs are above all the other NLDs. In particular, CT3 is 14 times larger than the other NLDs at $S_n$, with the differences increasing with $E^*$. The model by Hilaire et al. predicts the lowest level density.

As for the GSF, we employed two analytical descriptions, the model by Kopecky and Uhl (KU) [32] and the Simple Modified Lorentzian model (SMLO) by Goriely et al. [33]. In both cases we used the parameters given by Talys [26] for $^{208}$Pb. We also considered the results of Hartree-Fock-Bogolyubov (HFB) and Quasi-particle Random Phase Approximation (QRPA) calculations based on the Gogny D1M nuclear interaction [34], which we have denoted as D1M+QRPA.

We have combined the six descriptions for the NLD with the three models for the GSF leading to 18 calculations. The γ and the neutron emission probabilities are directly related to the number of final levels that are reached after the emission of the first γ ray of the cascade and after the emission of the neutron, respectively. In the $E^*$ range studied here, the neutron emission leads to the population of discrete low-lying levels of $^{207}$Pb and only the γ-emission is affected by the level density. We therefore expect that, for a given description of the GSF, the larger the level density the larger the γ emission probability and the smaller the neutron emission probability.

In [9], we observed strong discrepancies between the calculations and our results below $S_n$, which are due to the $E^*$ resolution, $\Delta E^*$. As is done in [9], to consider the effect of the $E^*$ resolution, we have convoluted the calculations with $\Delta E^*$. In the present case, the convolution was done with Gaussians with standard deviations varying as a function of $E^*$ as illustrated by the pink squares in Fig. 4 (a). To evaluate the degree of agreement between the calculations and our data we have computed the reduced Chi-squared $X^2$, defined as:

$$X^2 = \frac{1}{n}\sum_{i=1}^{n}\frac{\left(P^c_{\chi,i}-P_{\chi,i}\right)^2}{\left(\Delta P_{\chi,i}\right)^2} \tag{8}$$

where $P^c_{\chi,i}$ are the calculated probabilities and $n$ is the number of degrees of freedom. Note that in the present case, no adjustments have been made to the model parameters, we are not performing a fit. Consequently, the value of $n$ is equal to the number of data points, specifically $n = 28$, given that there are 14 data points from $P_n$ and 14 data points from $P_\gamma$. The $X^2$ values obtained for each calculation with and without convolution are given in figure 12. The $X^2$ values after convolution (panel a) are significantly lower than the ones without convolution (panel b). Since we have 28 degrees of freedom, calculations with reduced $X^2$ values above 1.72 can be rejected with a confidence level of 99%. Consequently, our data clearly rule out all the calculations without convolution. We obtain, the largest deviations before and after convolution with the CT3 and SMLO combination (see values at the intersection of columns 3 and rows 3 in figure 12). Our data indicate that the disagreement arises from the CT3 level density, as this level density leads to reduced $X^2$ values well above 1.72, independently of the GSF (see values in columns 3). Similarly, all the calculations that use the BSFG level density are excluded by our data. We obtained the smallest reduced $X^2$ value with the combination using the KU GSF model and the NLD by Hilaire et al. In fact, the level density calculated by Hilaire et al. gives the best agreement with our data for all the considered GSFs, demonstrating a preference of our data for the lower NLDs.

The calculations before and after convolution with the $E^*$ resolution which have the smallest and the largest reduced $X^2$ values are compared with our results from the first kinematic solution in Fig. 13. All the other combinations after convolution with reduced $X^2$ values below 1.72 are represented by the light blue shaded area. Regarding $P_n$, a comparison of the calculations using the KU GSF model and the microscopic NLD by Hilaire et al. before and after convolution (Fig. 13 (b)) reveals that the convoluted calculation is in much better agreement with our data below $S_n$. The structures of the calculation above $S_n$ are smoothed out after convolution, but the impact of the convolution is much weaker than below $S_n$. We observe a very good agreement between our experimental data and the convoluted calculation over the whole $E^*$ range covered by our data. In contrast, the convoluted calculation based on the CT3 and SMLO models clearly underestimates our results above 8 MeV. The comparison of the different calculations with our results for $P_\gamma$ shows the same tendencies as the ones observed for $P_n$. However, the conclusions are less clear due to the larger fluctuations and uncertainties of the $P_\gamma$ data. As expected, the calculations show that the effect of the level density on $P_\gamma$ goes in the opposite direction to $P_n$. The calculation using the NLD by Hilaire et al. gives the lowest $P_\gamma$ and the highest $P_n$, whereas the calculation based on the CT3 level density gives the highest $P_\gamma$ and the lowest $P_n$, see Fig. 13.

The results for $P_n$ obtained with the second kinematic solution presented in [9] also show that for a given model of the GSF, the CT3 and the BSFG level density models give the largest reduced $X^2$ values, and that the combination based on the CT3 and the SMLO models gives the largest reduced $X^2$ deviations of all the combinations. The calculation that most closely aligns with the data from the second kinematic solution uses the D1M+QRPA GSF model and the microscopic NLD by Goriely et al. This calculation is not excluded by the data from the first kinematic solution presented in this work, because it is in fact very similar to the calculation employing the KU GSF model and the microscopic NLD by Hilaire et al., as shown in Fig. 14 (a).

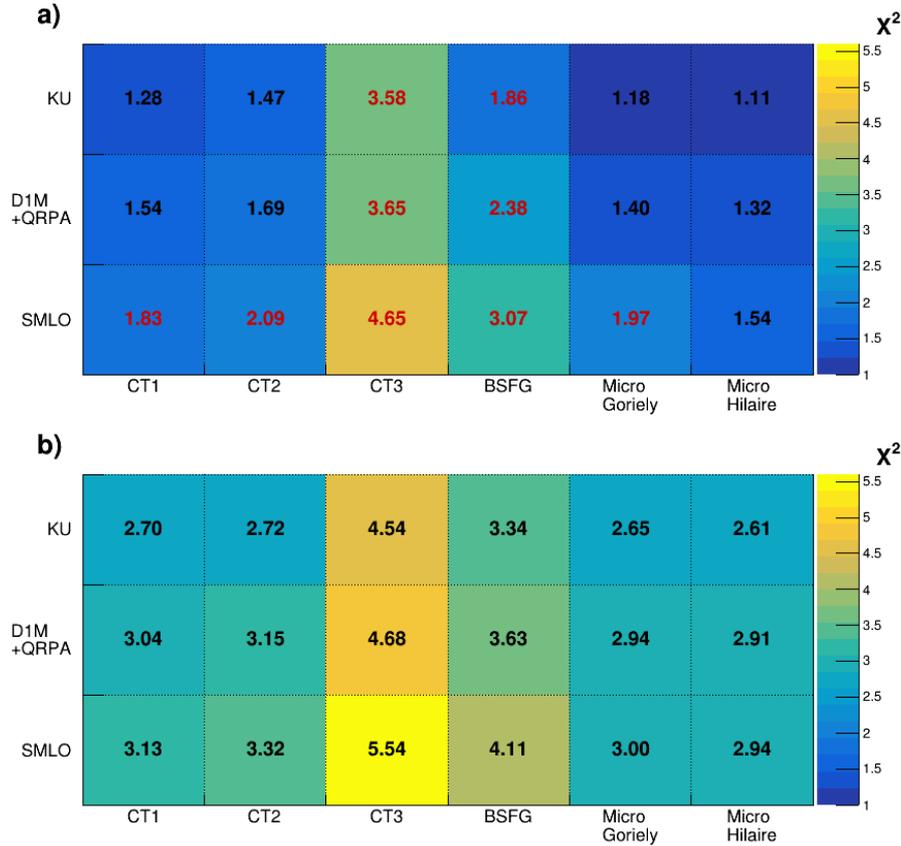

**Figure 12**: Reduced $X^2$ values obtained with the different combinations of the nuclear level densities (NLDs) and γ-strength functions (GSF). The NLDs are indicated in the horizontal axis and the GSFs in the vertical axis, see text for the details on the models. For each combination of GSF and NLD we give the value obtained with the calculation before convolution with the excitation energy resolution (panel b) and after convolution (panel a). The color scale is related to the value of the reduced $X^2$. The combinations with red values in panel (a) are excluded by our data with a confidence level of 99%.

Figure 14 (b) shows the calculations of Fig. 14 (a) after convolution with the excitation energy resolution. The convoluted calculation for the second kinematic solution is lower than the convoluted calculation for the first kinematic solution below 7.5 MeV, and becomes higher above 7.5 MeV. This effect is due to the differences between the excitation energy resolution $\Delta E^*$ for the first and second kinematic solutions, $\Delta E^*$ being significantly smaller for the second kinematic solution, see Fig. 4 (a). This $\Delta E^*$ effect may explain the differences we observed in Fig. 10 between the data for $P_n$ from the first and second kinematic solutions near $S_n$.

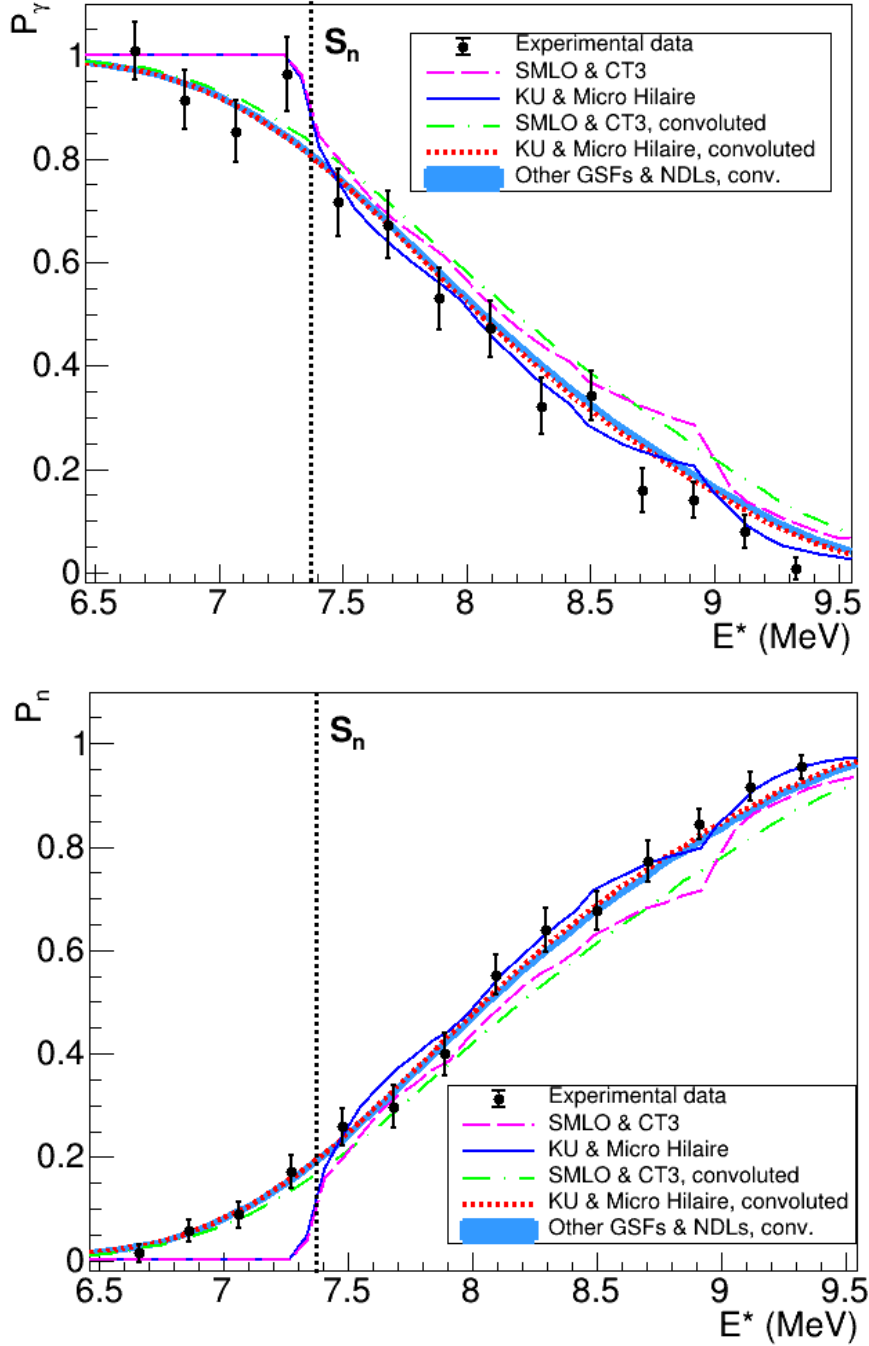

**Figure 13:** Probabilities for γ (a) and neutron (b) emission as a function of the excitation energy $E^*$ of $^{208}$Pb obtained with the data from the first kinematic solution compared with TALYS calculations, see text for details. The vertical lines indicate the neutron separation energy $S_n$ of $^{208}$Pb.

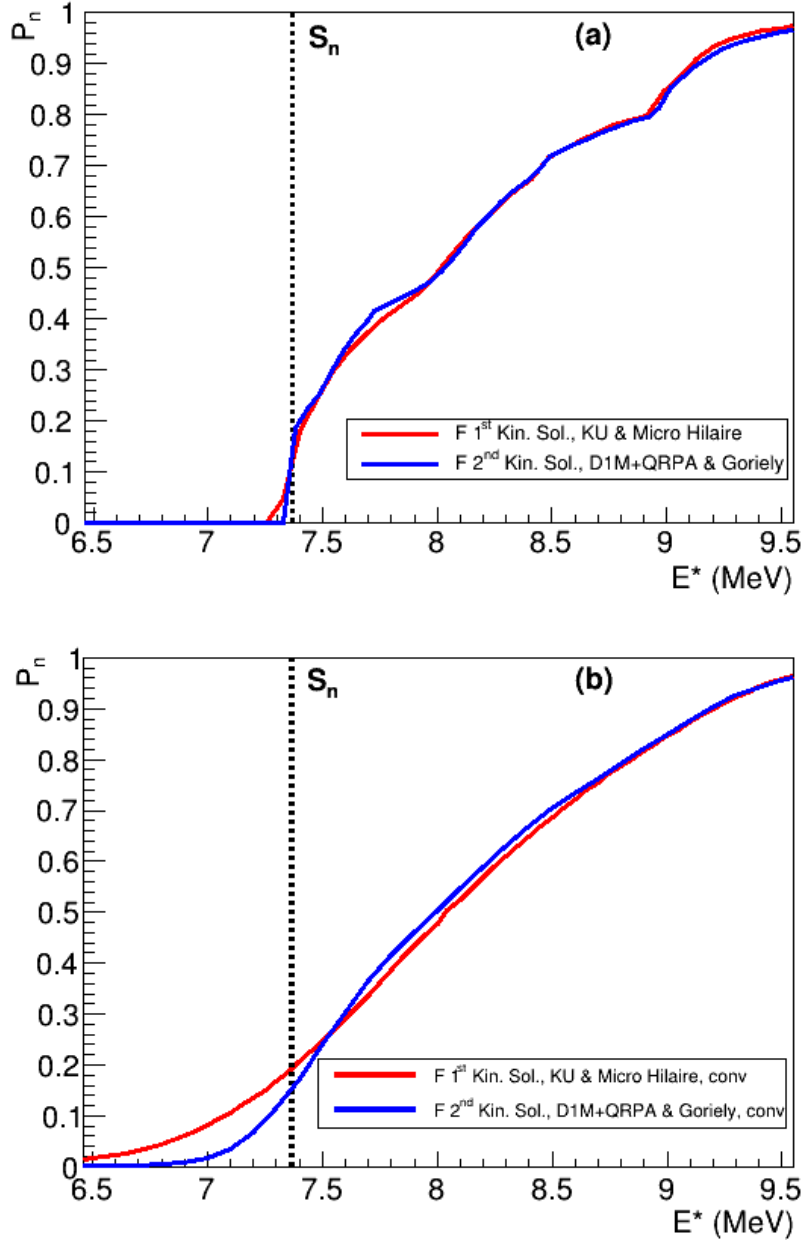

**Figure 14**: Calculations of the neutron emission probability as a function of the excitation energy of $^{208}$Pb leading to the smallest reduced $\chi^2$ deviations for the 1$^{st}$ and 2$^{nd}$ kinematic solutions before (panel a) and after (panel b) convolution with the excitation energy resolution. The calculations have been performed with the J$^\pi$ distributions corresponding to $\theta_{cm}$ range of the first and second kinematic solution. The vertical lines indicate the neutron separation energy $S_n$ of $^{208}$Pb.

## VII. Conclusions and perspectives

We have for the first time simultaneously measured the γ and neutron emission probabilities as a function of the excitation energy of $^{208}$Pb. The $^{208}$Pb nucleus was excited via the inelastic scattering $^{208}$Pb(p,p') reaction. The measurement was performed in inverse kinematics at the ESR storage ring of GSI/FAIR with a revolving $^{208}$Pb$^{82+}$ cooled beam at 30.77 MeV/nucleon repeatedly interacting with a windowless hydrogen gas-jet target of ultra-low areal density. The restoration of the beam quality after each passage through the target by the electron cooler of the ESR allowed us to minimize energy loss and straggling effects in the target, which have a significant impact on the excitation energy resolution. In this experiment. the excitation energy resolution ranged between 240 and 420 keV RMS and was dominated by the uncertainty in the scattering angle of the detected protons caused by the target radius of 2.5 mm. Thanks to the ultra-low density of the target, the heavy residues $^{208}$Pb and $^{207}$Pb produced after γ and neutron emission, respectively, came out of the target as bare ions. These residues were then fully separated by the ring dipole magnet located downstream from the target and were detected with outstanding efficiencies. The efficiency for the neutron emission channel was 100% and the efficiency for the γ-emission channel varied between 33 and 100%. This shows the considerable advantage of our new technique over standard experiments in direct kinematics and single-pass experiments in inverse kinematics, where the detection efficiencies are much lower. Thanks to the 100% and precise detection efficiency for the neutron-emission decay channel, we were able to achieve a relative uncertainty for the neutron-emission probability of only 3%. In the range of excitation energy covered by our data the only two open decay channels are γ and neutron emission. Thus, the γ and neutron-emission probabilities must add up to one. This is fulfilled by our results and validates our new technique.

The comparison of our results for the neutron emission probability with TALYS calculations allowed us to test different models for the nuclear level density and γ-ray strength function of $^{208}$Pb available in the literature. Our data rule out the level density descriptions of [25] and [28]. The two latter level densities are significantly larger than the other tested nuclear level density descriptions [26, 24, 30, 31]. Our results are best described with the microscopic level density description of Hilaire et al. [31] and the γ-ray strength function by Kopecky and Uhl [32].

In the future, we will add a fission detector to our setup, increase the solid angle of the target residue detector and use a target with a thinner diameter. Our simulations show that with a target radius of 0.5 mm we will be able to achieve an excitation energy resolution of about 100 keV, which will allow us to precisely measure the strong dependence of the decay probabilities on excitation energy at the particle and fission thresholds. With these improvements we will be able to measure simultaneously and with high precision the probabilities for fission, γ emission, one neutron and even two neutron emission of many short-lived nuclei of interest in astrophysics and applications, which are available as radioactive ion beams.

**Acknowledgements**

This work is supported by the European Research Council (ERC) under the European Union's Horizon 2020 research and innovation programme (ERC-Advanced grant NECTAR, grant agreement No 884715). The results presented here are based on the experiment E146, which was performed at the ESR storage ring of the GSI Helmholtzzentrum fuer Schwerionenforschung, Darmstadt (Germany) in the context of FAIR Phase-0. We thank the Prime 80 program from the CNRS for funding the PhD thesis of MS and the GSI/IN2P3 collaboration 19-80. JG, YuAL, RR and ThS acknowledge support by the State of Hesse within the Cluster Project ELEMENTS (Project ID 500/10.006). AH is grateful for funding from the Knut and Alice Wallenberg Foundation under KAW 2020.0076.